# Personalized Motion Guidance Framework for Athlete-Centric Coaching

Ryota Takamido[a*], Chiharu Suzuki[a], Hiroki Nakamoto[b]

[a]Sports Innovation Organization, National Institute of Fitness and Sports in Kanoya, Kanoya, Kagoshima 891-2393, Japan

[b]Faculty of Physical Education, National Institute of Fitness and Sports in Kanoya, Kanoya, Kagoshima 891-2393, Japan

Corresponding author: Ryota Takamido (Email: rtakamido@nifs-k.ac.jp)

## Abstract

**Background:** A major challenge in sports science is bridging the gap between controlled, group-level findings and the real-world need for individualized coaching. Recent advances in machine learning—particularly in latent space feature manipulation—offer the potential to generate and provide feedback on improved motion patterns, based on each athlete's original motion data.

**Methods:** This study newly developed a Personalized Motion Guidance Framework (PMGF) to enhance athletic performance by generating individualized motion-refinement guides using generative artificial intelligence techniques. PMGF leverages a variational autoencoder to encode motion sequences into athlete-specific latent representations, which can then be directly manipulated to generate meaningful guidance motions. Two manipulation strategies were explored: (1) smooth interpolation between the learner's motion and a target (e.g., expert) motion to facilitate observational learning, and (2) shifting the motion pattern in an optimal direction in the latent space using a local optimization technique. As these manipulations are applied to individual-specific latent representations, the generated guidance motions vary among individuals.

**Results:** The results of the validation experiment with data from 51 baseball pitchers revealed that (1) PMGF successfully generated smooth transitions in motion patterns between individuals across all 1,275 pitcher pairs, and (2) the features significantly altered through PMGF manipulations reflected known performance-enhancing characteristics, such as increased stride length and knee extension associated with higher ball velocity, indicating that PMGF induces biomechanically plausible improvements.

**Conclusion:** The generated outputs can support personalized, athlete-centric coaching practices, such as delivering more individualized visual feedback to enhance observational learning. Although PMGF has limitations to be addressed in future work, it shows the potential of generative AI as a practical tool to bridge group-level findings and individualized coaching needs.

**1. Background**

Traditional sports performance analysis has primarily focused on identifying the typical kinematic and kinetic characteristics of athletes, either within specific groups (e.g., experts) or between groups (e.g., experts versus novices) [1]. Skilled athletes within the same sport tend to exhibit many shared characteristics, as they adapt to common task constraints such as rules and equipment [2,3]. These insights into the common characteristics of athletes can help guide the development of effective coaching strategies in many sports [4].

However, a major challenge in contemporary sports science is personalizing these insights for athletes with unique bodies and movement patterns. For example, in baseball pitching, previous studies have demonstrated a significant correlation between the knee extension angle during the throwing arm acceleration phase and the pitch velocity [5]. However, because individual pitchers exhibit distinct movement patterns, improving the knee extension angle may require different refinements for each athlete. One pitcher may need to refine the trunk tilt angle during weight transfer, whereas another may need to adjust the position of the leading foot. In this case, rather than providing direct instructions such as 'extend the knee at ball release,' it may be more effective to offer personalized guidance based on each pitcher's characteristics. Currently, such personalization relies on expert coaches' knowledge and practical experience (Figure 1a). The importance of analysis and feedback tailored to individual characteristics has been increasingly emphasized [1]. Therefore, bridging the gap between identifying shared characteristics across athletes and delivering personalized athlete-centric coaching is a key challenge in contemporary sports science.

This study was conducted to address this issue by developing a *Personalized Motion Guidance Framework (PMGF)*, which enhances sports performance by generating individualized motion refinement guides using generative artificial intelligence (AI) techniques (Figure 1b). Recent advancements in motion generation techniques, such as *motion style transfer* and *motion synthesis*, have leveraged deep generative models to refine movement patterns within a low-dimensional latent space [6,7]. These latent spaces, learned through machine learning models such as Variational Autoencoders (VAEs) [8], preserve semantic continuity and allow smooth and interpretable transitions between different movement patterns. Building on these techniques, the PMGF manipulates an athlete's motion within the latent space to generate improved version of their current movement patterns. Specifically, PMGF explores two manipulations: (a) transforming an athlete's motion style toward that of a target expert and (b) identifying and shifting toward the optimal neighboring motion that maximizes key biomechanical features reported in previous performance analysis studies. These manipulations provide personalized guidance for each athlete, either to imitate an ideal motion or to improve specific biomechanical features in harmony with inherent movement patterns. Therefore, this approach addresses a key limitation in current sports science: the challenge of translating group-level findings into personalized feedback.

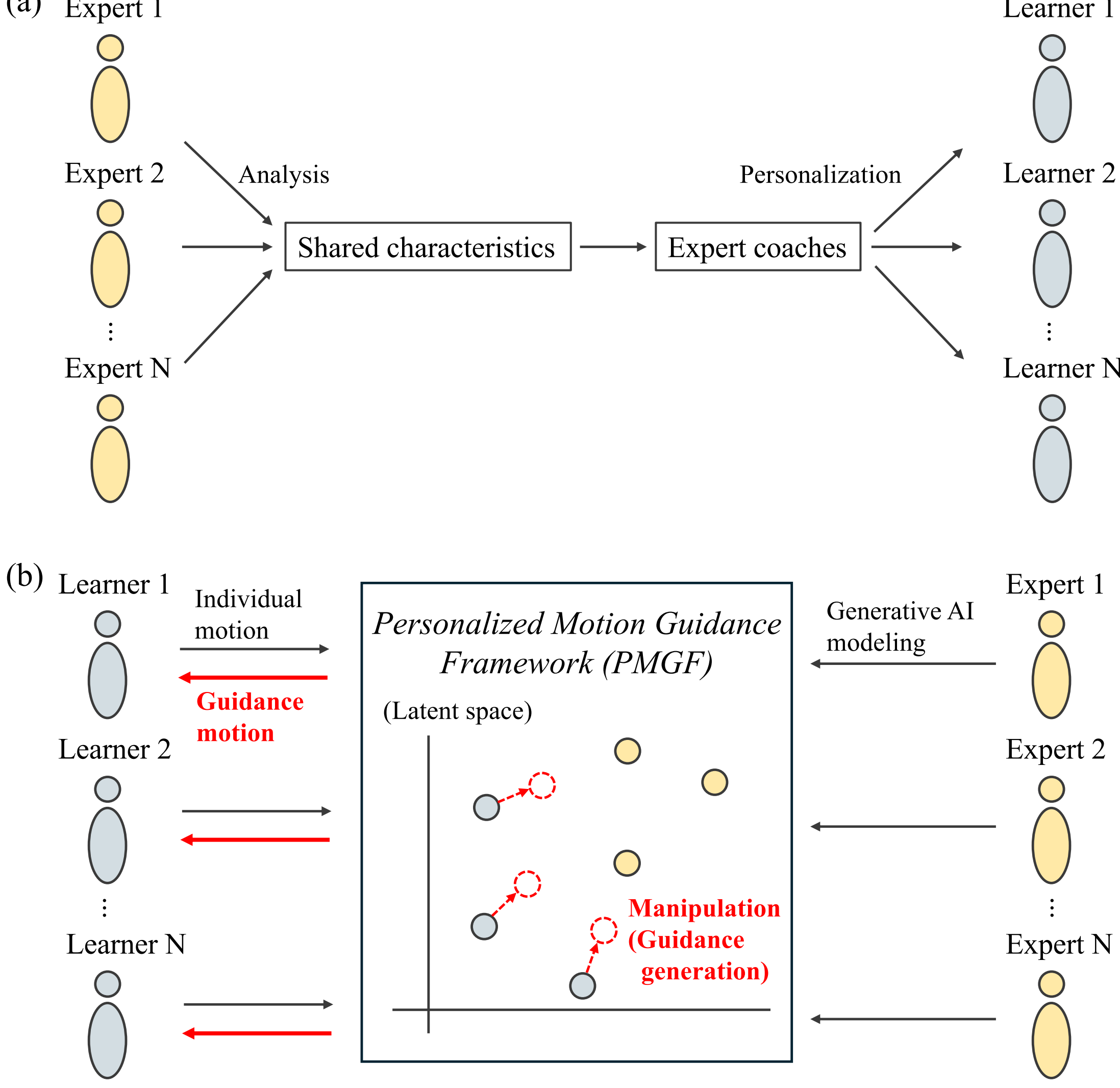

**Fig. 1 Conceptual comparison between conventional performance analysis and PMGF** (a) Traditional approach that extracts shared characteristics. (b) PMGF approach that models and manipulates individual motion patterns to generate personalized motion guidance. PMGF, Personalized Motion Guidance Framework

## 2. Methods

This study aims to develop a generative AI–driven framework, the *PMGF*, designed to provide personalized motion refinement aligned with each athlete's unique movement characteristics through latent space representations, thereby enhancing sports performance in an athlete-centric manner. The PMGF source code, anonymized dataset, and analysis scripts used in this study are publicly available on GitHub (https://github.com/takamido/PMGF).

### 2.1 Overview of PMGF

Figure 2 illustrates the overall architecture and workflow of the PMGF using baseball pitching as an example. Although this study focuses on the development and evaluation of the PMGF using baseball pitching as the target movement—because of the complexity of its pattern and the high level of skill required for mastery—the framework is theoretically applicable to a wide range of sports skills, regardless of the type of motion.

The PMGF consists of two primary components: (1) a Variational Autoencoder (VAE) with transformer-based encoder and decoder architectures [9] for learning motion representations, and (2) a manipulation module that performs transformations in the learned latent space either toward a target expert's motion or in a direction that refines biomechanical features such as joint angles. In the development process, the PMGF first trains a Transformer-VAE to reconstruct each athlete's motion in the dataset (Figure 2a). The key feature of this approach is that the VAE provides a semantically smooth latent space, enabling continuous and natural transformations of the generated motion sequences through latent vector manipulation [10]. Intuitively, this enables perturbations in the latent space to be mapped to meaningful changes in motion patterns in real space, such as a slight increase in elbow angle. The trained Transformer-VAE mapped each athlete's movement pattern to the corresponding position in the latent space (Figure 2b).

Subsequently, the PMGF utilizes the constructed latent space to generate individually refined motions through two types of manipulation, which are then presented as guidance for athletes. The first manipulation transformed the learner's motion style toward that of another player exhibiting an ideal movement pattern for the learner (Figure 2c). This approach is based on the framework of *observational learning*, which facilitates motor learning by observing movements [11,12]. From the perspective of observational learning, the primary advantage of the PMGF lies in its ability to demonstrate 'intermediate' motions between the learner's own movements and those of others. In previous research, various demonstration models have been used, including the self-model, which presents the learner's own movements; the expert model, which presents those of skilled performers; and the coping model, which illustrates how a novice improves over time. The strengths and limitations of each method have been discussed [12]. In this context, the transferred motion generated by the PMGF can be positioned as a model that combines the strengths of both self- and expert-models—an expert-like motion that remains closely aligned with the learner's own movements. This approach offers more sophisticated demonstrations for the expert–self combining model, a concept that has gained increasing attention in recent research [13-15]. It can also be positioned as a 'self-coping' model that illustrates a learner's own process of improvement.

Furthermore, the second manipulation adjusted the learner's motion to maximize key biomechanical features, as defined by previous studies (Figure 2d). While the former manipulation involved a global transformation of the entire motion pattern, this approach primarily targeted the critical features that influence performance. Specifically, we identified the optimal motion within the neighborhood of a given pitcher's latent representation that maximizes biomechanical features related to ball velocity, such as knee extension angle and forward trunk tilt [16,17], using the Evolution Strategy (ES) algorithm [18], a bio-inspired optimization technique. This approach enabled the generation of guidance based on the athlete's original movements while incorporating improvements in key biomechanical features.

In summary, the PMGF leverages latent space manipulation to generate individualized motion refinements through two distinct strategies: style imitation and biomechanical feature-oriented refinement, tailored to each athlete's unique movement pattern. From a technical standpoint, existing motion-generation studies have primarily focused on producing representative motions for specific categories or conditions (e.g., walking) [19]. In contrast, this study represents a novel attempt to apply motion-generation techniques to motor learning, where each athlete's unique movement pattern plays a critical role. The following section provides a detailed technical explanation of each component of the PMGF.

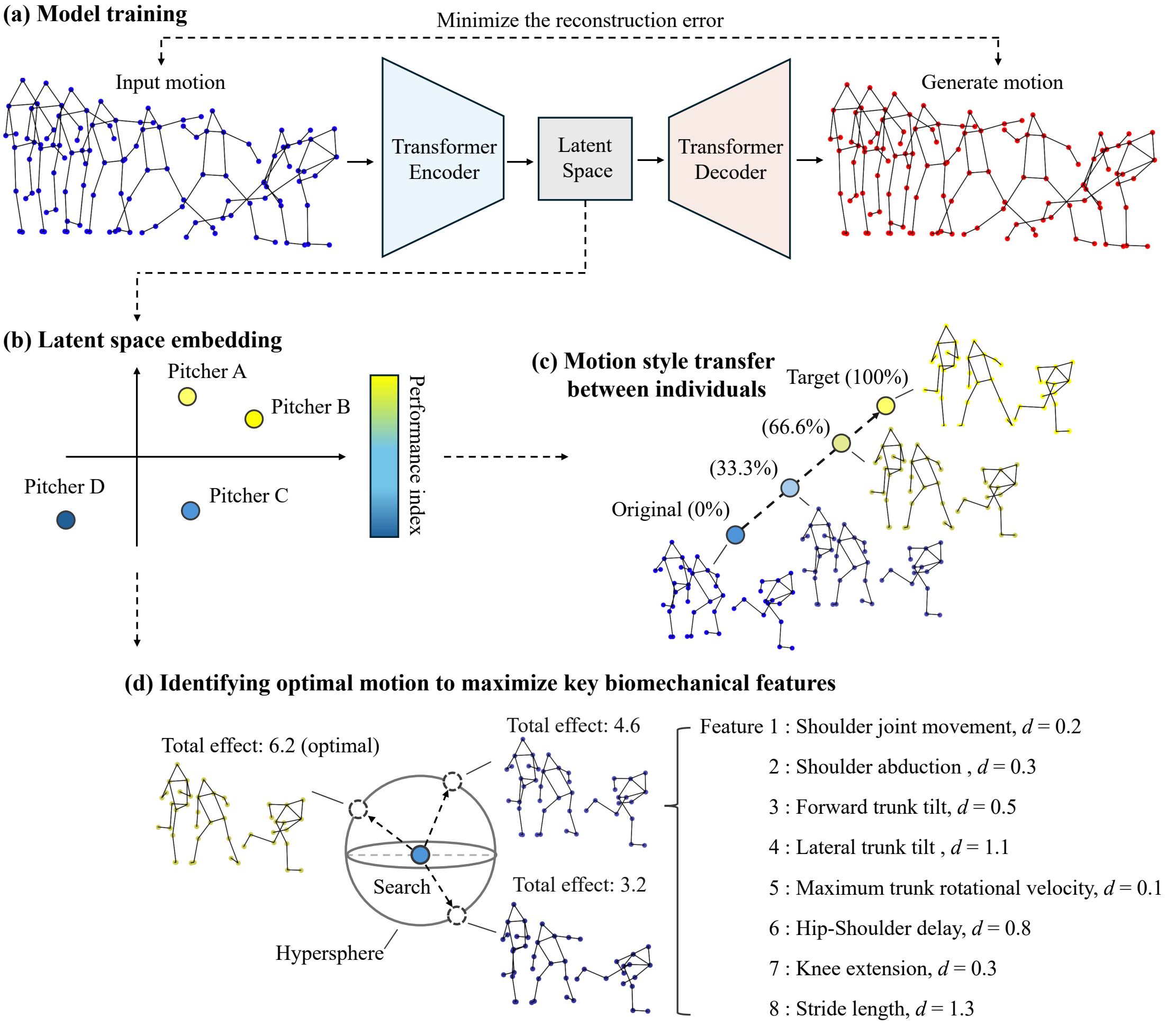

**Fig. 2 Overview of the PMGF workflow using baseball pitching as an example.**

### 2.2 Transformer-VAE for motion representation and reconstruction

The latent representations of athletic motion sequences were learned using a Transformer-based VAE (Figure 2a). In recent years, transformer architectures have emerged as powerful tools across a wide range of machine learning applications, including image processing [20] and robot motion generation [21]. One key advantage is their ability to model long-range temporal dependencies and interactions [22], which are particularly important for capturing the dynamics of athletic movements. For example, in baseball pitching, posture during the weight transfer phase can influence posture at ball release, making it essential to use a model capable of capturing such temporally distant relationships. Transformers possess these strengths and have therefore been successfully applied in several studies on human motion generation [23,24], demonstrating their potential to capture the complex dynamics of full-body movements. Leveraging these capabilities, the PMGF integrates a transformer into the encoder and decoder components of the VAE to enhance motion representation quality.

The input to the encoder consists of a motion sequence of length $T$, where each frame contains the three-dimensional positions of $n$ joints. Thus, the input tensor is represented as a sequence $\boldsymbol{x} \in \mathbb{R}^{T \times n \times 3}$. The input tensor was flattened and linearly projected onto a model space with fixed dimensionality $d$. Sinusoidal positional encodings were added to the frame embeddings, which were then passed through a transformer encoder consisting of multihead self-attention and feed-forward layers. As this study focuses primarily on applying the transformer architecture to motion generation, detailed descriptions of its internal components are omitted. Further details are available in the original implementation [9].

After processing by the encoder, the output embeddings are aggregated and transformed through two linear projections to obtain the parameters of the latent vector $\boldsymbol{z}$: a mean vector $\mu \in \mathbb{R}^{d_z}$ and a log-variance vector $log\sigma^2 \in \mathbb{R}^{d_z}$, where $d_z$ represents the dimensionality of the latent space. The decoder receives the latent vector $\boldsymbol{z}$ and transforms it into a temporal sequence using a transformer decoder architecture. Positional encodings are incorporated again to reintroduce temporal structure during decoding. The output of the decoder is linearly projected back to the original motion space, yielding a reconstructed sequence of the input human motion $\hat{\boldsymbol{x}} = \mathbb{R}^{T\times n\times 3}$.

The following loss function is typically used to train a VAE [8]:

$$\mathcal{L} = \mathcal{L}_{recon} + \lambda_{KL}\mathcal{L}_{KL}, \tag{1}$$

where $\mathcal{L}_{recon}$ denotes the reconstruction loss, typically computed as the mean squared error (MSE) between the input motion sequence and its reconstruction. $\mathcal{L}_{KL}$ represents the Kullback–Leibler divergence [25] between the approximate posterior $q(\boldsymbol{z}|\boldsymbol{x})$ and a standard normal prior $p(\boldsymbol{z})$, encouraging the latent space to be continuous and smooth. $\lambda_{KL}$ is a hyperparameter that controls the relative importance of the $\mathcal{L}_{KL}$. We also added a motion-speed penalty term related to the motion speed with reference to previous research [26] to ensure that the generated motion sequences accurately reflected the athlete’s movements:

$$\mathcal{L} = \mathcal{L}_{recon} + \lambda_{KL}\mathcal{L}_{KL} + \lambda_{speed}\mathcal{L}_{speed}, \quad (2)$$

where $\mathcal{L}_{speed}$ is the penalty term that aligns the dynamic characteristics of the reconstructed motion with those of the original, and $\lambda_{speed}$ is the hyperparameter controls the relative importance of $\mathcal{L}_{speed}$. This term penalizes discrepancies in overall movement speed, defined as the mean squared error between the joint-wise velocity magnitudes of the original and reconstructed motions. Designing loss functions that incorporate kinematic characteristics can improve the fidelity of generated human motions.

### 2.3 Latent space manipulation for motion guide generation

Each athlete's motion was mapped to a corresponding point in the latent space using the trained Transformer-VAE described above (Figure 2b). Here, due to the smooth and continuous nature of the VAE latent space, a slightly modified latent representation $\boldsymbol{z}'$, derived from an athlete's original representation can be decoded to produce a motion that reflects a subtle but meaningful variation in the athlete's movement pattern.

In particular, the PMGF implements two manipulation strategies to generate guide motions for athletes and coaches. The first strategy transforms the motion style of the learner, typically represented by the input data, into a designated target motion (Figure 2c). Examples of target motions include those performed by more skilled athletes or by learners themselves during their best performance. This manipulation is represented by the following equation:

$$\boldsymbol{z}' = (1-\alpha)\boldsymbol{z}_{original} + \alpha\boldsymbol{z}_{target}, \quad (3)$$

where $\alpha \in [0,1]$ is the interpolation parameter that controls the degree of transformation from the original to the target motion pattern, gradually increasing the value of $\alpha$ allows the learner's motion pattern to be smoothly transformed toward the target. As described in the previous section, the generated motion serves as a hybrid demonstration that combines the strengths of self and expert models and functions as a self-coping model rooted in the learner's own movement.

The second strategy shifts the learner's own motion to optimize the biomechanical features related to ball velocity using a local optimization technique (Figure 2d). Specifically, this study introduced an Evolution Strategy (ES) [18] to identify an optimal motion in the neighborhood of the athlete's latent representation, constrained to lie on the surface of an *r*-radius hypersphere. ES is a population-based gradient-free optimization method that iteratively refines the search direction by evaluating a set of perturbations applied to candidate solutions. It is particularly suitable for continuous black-box optimization, where the relationship between inputs and outputs is unknown, as in the present case between the latent-space representation and biomechanical features [27]. Given $\boldsymbol{u}$ is the population of direction vectors; the manipulation is represented by the following equation:

$$\boldsymbol{z}' = \boldsymbol{z}_{original} + r\boldsymbol{u}. \tag{4}$$

In Equation (4), the size of the manipulation is controlled by the value of the radius $r$. Intuitively, this corresponds to sampling candidate shift directions from the surface of an $r$-radius hypersphere. The manipulated latent representations $\boldsymbol{z}'$ are decoded by the Transformer-VAE to generate corresponding pitching motions, which are then evaluated using a biomechanical fitness function that reflects improvements in key features. In this study, the fitness function $f(\boldsymbol{u})$ is specifically defined using a Nash product formulation as follows:

$$f(\boldsymbol{u}) = \prod_{i=1}^{K} (1 + \alpha w_i \Delta_i)^{1/K}, \tag{5}$$

where $K$ represents the number of feature variables, and $\alpha$ is a scaling parameter controlling the sensitivity of the aggregation. The Nash product was adopted to prevent the optimization from being dominated by an extreme improvement in a single variable and to promote a balanced enhancement across all biomechanical features. $\Delta_i$ represents the effect of the manipulation on the $i$-th feature, defined as follows:

$$\Delta_i = \frac{f_{rec,i} - f_{ori,i}}{d_i}, \tag{6}$$

where $f_{rec,i}$ and $f_{ori,i}$ represent the values of the $i$-th feature before and after manipulation, respectively. The normalization term $d_i$ was manually specified to account for the differences in units and scales across the features. The direction vector $\boldsymbol{u}$ is iteratively updated toward candidates with higher fitness values based on Gaussian perturbations $\boldsymbol{\sigma}$ sampled around the current mean direction. After multiple iterations, the algorithm converges to an optimal latent space direction, representing the most beneficial transformation of the athlete's pitching motion pattern.

**2.4 Dataset**

We evaluated the proposed PMGF framework using a dataset of baseball pitching motions performed by athletes with varying skill levels. The dataset consisted of baseball pitching motions and corresponding ball velocity data collected from 51 athletes. The participants included four high school players, 10 collegiate athletes, 21 industrial league players, three independent league players, and 13 professional league players. For each athlete, five fastball pitches were recorded using 16 optical motion capture cameras (Raptor-E, Motion Analysis Corporation, Santa Rosa, CA, USA) at a sampling rate of 200–500 Hz. The positions of 15 anatomical landmarks were extracted from the measured data, including the parietalis (head), bilateral acromion (shoulder), lateral epicondyle of the humerus (elbow), radial styloid process (wrist), greater trochanter of the femur (hip), lateral condyle of the femur (knee), and the heel and top of the shoes. The ball velocity (initial speed) of each pitch was also recorded as the performance index using the TrackMan Baseball system (TrackMan, Vedbæk, Denmark). The mean ball velocity and standard deviation of the 51 pitchers was 82.61 ± 5.37 mph (range: 68.47−94.17 mph).

**2.5 Preprocessing**

Several preprocessing steps were applied to the collected motion and velocity data to prepare the input data for PMGF. A fourth-order Butterworth filter was applied to the raw motion data, with an optimal cutoff frequency adjusted according to the method proposed in [28]. Subsequently, the ball release time for each pitch was identified as the moment when the wrist velocity of the throwing arm reached its maximum value. From this event, a fixed-duration segment was extracted from each motion sequence, spanning from 1.0 seconds before to 0.2 seconds after ball release, covering the weight-shift, release, and follow-through phases. The 3D positions of the 15 joints were extracted from this segment and temporally normalized to 101 frames. Finally, the joint position data were standardized to a normal distribution across the entire dataset. The ball velocity data for each pitch were standardized across all pitchers.

**2.6 VAE model training**

The PMGF was trained based on the developed dataset. For the Transformer-VAE, both the encoder and decoder used transformer modules consisting of three layers with eight attention heads and a model dimension of 256. The dimensionality of the latent space $d_z$ was set to 256. The weights for each penalty term in the loss function (Equation 2) were set as $\lambda_{KL} = 1 \times 10^{-3}$ and $\lambda_{speed} = 1.0$. It is worth noting that the value of $\lambda_{KL}$ used in this study is substantially smaller than commonly adopted values (typically around 0.01~0.1). This choice was made because larger values of $\lambda_{KL}$ resulted in degraded reconstruction quality for high-speed, high-acceleration movements such as pitching. For the same reason, the weight of the speed penalty $\lambda_{speed}$ was set relatively large to emphasize biomechanical fidelity. Given these hyperparameter settings, the Transformer-VAE was trained using the Adam optimizer with a learning rate of $1 \times 10^{-4}$. The model was trained for 2000 epochs using all motion samples (255 pitches in total). The batch size was set to 64. Training was conducted on the Google Colab platform using an NVIDIA T4 GPU, and each training run required approximately 30–40 minutes. The performance of the Transformer-VAE was evaluated by aggregating reconstruction errors (root mean squared error [RMSE]) after training each joint across all pitchers. Only the mean vector was decoded during reconstruction to ensure reproducibility.

## 2.7 Analysis and evaluations

### 2.7.1. Motion style transfer between individuals

We evaluated whether the motion style transitions between athletes occurred in a smooth and continuous manner—that is, whether the generated motion sequences gradually approached the target motion, by conducting a style transfer analysis using the Dynamic Time Warping (DTW) technique (Figure 3) [29]. DTW is a fundamental method for measuring the similarity between two time series data and was used in a previous study to analyze the similarity of weight-shift patterns between two athletes [30].

Specifically, for each pitcher in the dataset, a representative motion sample (i.e., the fastest pitch) was selected as the reference. Using the trained Transformer-VAE, we extracted the latent representations $\boldsymbol{z}_{original}$ and $\boldsymbol{z}_{target}$ corresponding to the source (learner) and target (reference) players, respectively. The interpolation coefficient $\alpha$ was varied from 0 to 1 in increments of 0.1, yielding 10 evenly spaced interpolation steps. The intermediate latent vectors $\boldsymbol{z}'$ obtained at each step from Equation (3) were decoded using the trained VAE decoder, resulting in a sequence of eight intermediate motion trajectories positioned between the source and target styles.

The transition behavior of the generated motions was assessed by calculating DTW-based similarity scores ($s_{motion}$) between each interpolated motion and the two reference motions (original and target).

$$s_{motion} = \frac{\sum_{j=1}^{15} \sum_{c=1}^{3} DTW\left(m_{j,c}^{(1)}, m_{l,c}^{(2)}\right)}{101 \times 45}, \quad (7)$$

where $m_{j,c}^{(1)}$ and $m_{l,c}^{(2)}$ represent the temporal trajectories of the *j*-th joint along coordinate axis $c \in \{x, y, z\}$ for the two motions being compared. The final similarity score was normalized to the total number of joint axis pairs and time steps in the postprocessing stage. These scores quantitatively described how the generated motions gradually diverged from the original motion and approached the target motion as $\alpha$ increased. This analysis was repeated across all possible source–target athlete pairs (i.e., $\frac{51 \times 50}{2} = 1275$ unique combinations), providing a comprehensive assessment of the PMGF's latent space interpolation capability.

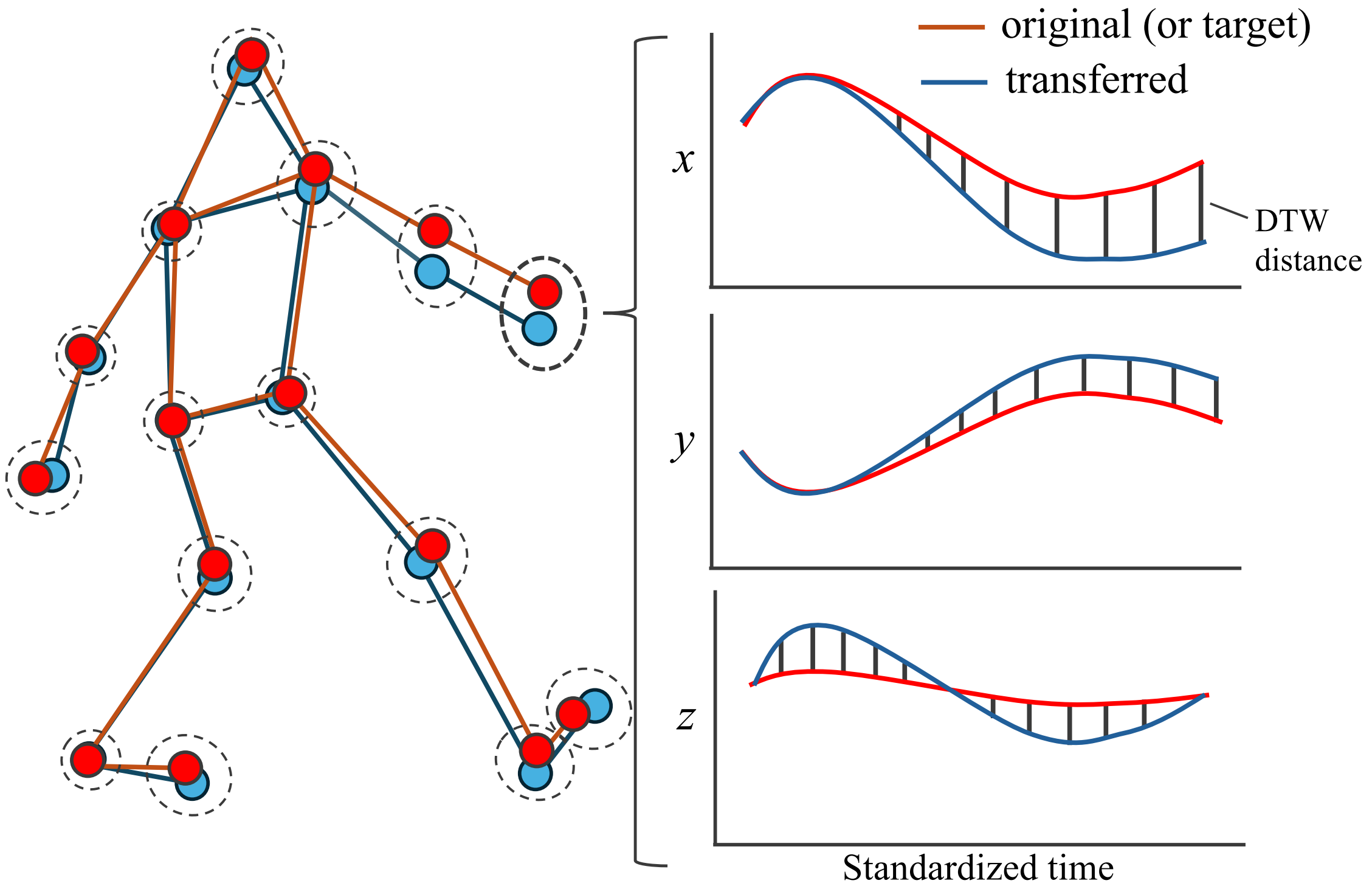

**Fig. 3 Schematic image of the DTW analysis for evaluating the similarity between two motion sequences**. The similarity metric defined in Equation (7) was calculated by summing the DTW distances computed for each joint illustrated in the figure. DTW, Dynamic Time Warping

#### 2.7.2. Identifying and shifting the optimal motion to maximize key biomechanical features

The capability of PMGF to generate improved motion patterns was verified by shifting a learner's latent representation in an optimal direction—on the surface of an $r$-radius hypersphere—to maximize biomechanical features. We conducted a biomechanical analysis of the generated motions and evaluated their consistency with findings reported in previous studies [5,16,17,31-36]. Specifically, we extracted eight biomechanical features empirically linked to pitching performance (Figure 4), as

described below:

(F1) *Shoulder joint movement* [31]: defined as the displacement of the non-throwing shoulder during the arm acceleration phase. This displacement was computed over the 10 times steps preceding ball release. Smaller values indicated more effective mechanics and were associated with higher ball velocities.

(F2) *Shoulder abduction* [32]: approximated as the angle between the vector from the throwing shoulder to the hip and from the shoulder to the elbow. The average angle was calculated over 10 frames before ball release. Although excessively large values may negatively affect performance [37], in this study increases in this angle were treated as positive effects to simplify the reward function.

(F3) *Forward trunk tilt* [16]: defined as the angle between the horizontal plane (ground) and the vector connecting the midpoint of the hips and shoulders at the moment of ball release. A larger forward tilt was positively associated with increased ball velocity.

(F4) *Lateral trunk tilt* [17]: defined as the lateral inclination of the torso, approximated by the angle of the vector connecting the stride-leg heel to the head projected toward the non-throwing arm side. A greater tilt was linked to higher pitch speeds.

(F5) *Maximum trunk rotational velocity* [33]: defined as the peak angular velocity of the trunk during torso rotation, was approximated using the angular velocity of the hips in the horizontal plane. Greater velocity was linked to higher pitch speeds.

(F6) *Hip–shoulder delay* [34]: quantified as the time difference between the peak angular velocities of the hip and shoulder in the horizontal rotational direction. This hip-shoulder delay reflected the degree of kinetic chain utilization and was positively correlated with ball velocity.

(F7) *Knee extension* [5]: calculated as the angle formed by the vectors from the lead knee to the hip and from the knee to the heel during ball release. Greater knee extension during the arm acceleration phase was associated with higher pitch velocity.

(F8) *Stride length* [35]: defined as the distance from the pivot foot heel in the initial frame (set to time 0) to the stride foot heel at the moment of ball release. Greater stride length is generally associated with higher pitch speeds.

It should be noted that the present study relied solely on 3D joint positional data. Therefore, certain variables, such as foot contact timing, must be approximated (e.g., by assuming that foot contact occurs 10 frames before ball release).

These features were optimized using the following procedure: first, we selected 17 of the 51 pitchers whose average ball velocity was ranked in the lower third. The average ball velocity of this group was 76.64 ±3.23 mph. Subsequently, the ES algorithm was applied to identify the optimal motion of the 17 pitchers, maximizing the fitness function composed of the eight biomechanical features described above (Equation 5). Optimization was performed independently for each athlete using five pitching trials per individual. In each iteration, 128 random perturbations were generated and mirrored to form a symmetric population of 256 candidate directions, whose fitness values were then computed to update the mean search direction toward higher-performing candidates. The search process was repeated for 20 iterations, with the search radius set to $r$ = 3.0. Although a larger radius may result in greater improvements in biomechanical features, it can also lead to larger deviations from the athlete's original motion. Therefore, to balance biomechanical enhancement with preservation of the original motion, the radius was set to $r$ = 3.0 in this study. The standard deviation of the Gaussian perturbations was fixed at $\boldsymbol{\sigma}$ = 0.1, and the learning rate for updating the search direction was *lr* =0.5. Biomechanical features were weighted $w_i = [-1, 1, 1, 1, 1, 1, 1,1]$, where the negative weight for shoulder joint movement ensured that smaller values were considered favorable. The sensitivity parameter of the Nash aggregation was set to $\alpha$ = 5.0. The normalization term $d_i$ in Equation (6) was manually specified according to the physical nature of each biomechanical feature to ensure comparability in scale. First, non-angular features (F1, F5, and F8) were normalized by their original magnitudes before shifting, with a small constant ($10^{-6}$) added to ensure numerical stability. Angular features (F2, F3, F4, and F7) were normalized by π. The timing feature (F6) was normalized by a fixed scale of 10 frames.

The resulting optimal direction vector $\boldsymbol{u}^*$ defined the most beneficial modification of each athlete's latent representation, which was then decoded to reconstruct an optimized motion pattern for subsequent biomechanical and qualitative analyses. For each pitcher, the mean value of each variable was computed across the five original and optimized motions. The group-level means across the 17 athletes were calculated for both the original and manipulated conditions. If the latent space manipulation introduced by the PMGF was biomechanically meaningful, the generated motions were expected to exhibit characteristic features aligned with existing studies.

We conducted paired *t*-tests for each of the eight biomechanical features to statistically evaluate the changes induced by manipulation. The significance threshold was set at 5%, and the Holm–Bonferroni method was applied to correct for multiple comparisons. Cohen's *d* was calculated to estimate the effect size of each manipulation-induced change. Given that the training process included sources of stochasticity, such as mini-batch shuffling and dropout layers, we further assessed model stability. For the purpose of this study, we trained and evaluated five independent instances of the PMGF model under identical conditions and performed biomechanical analysis for each instance. Finally, to examine the effect of the radius $r$, we performed optimization using $r$ = 1, 2, 3, 4 and 5 for each of the five model instances and computed the total effect size across the eight features for each value. The optimization was performed in the same Python environment used for training the VAE, while statistical tests were conducted in MATLAB (MATLAB R2022b, MathWorks).

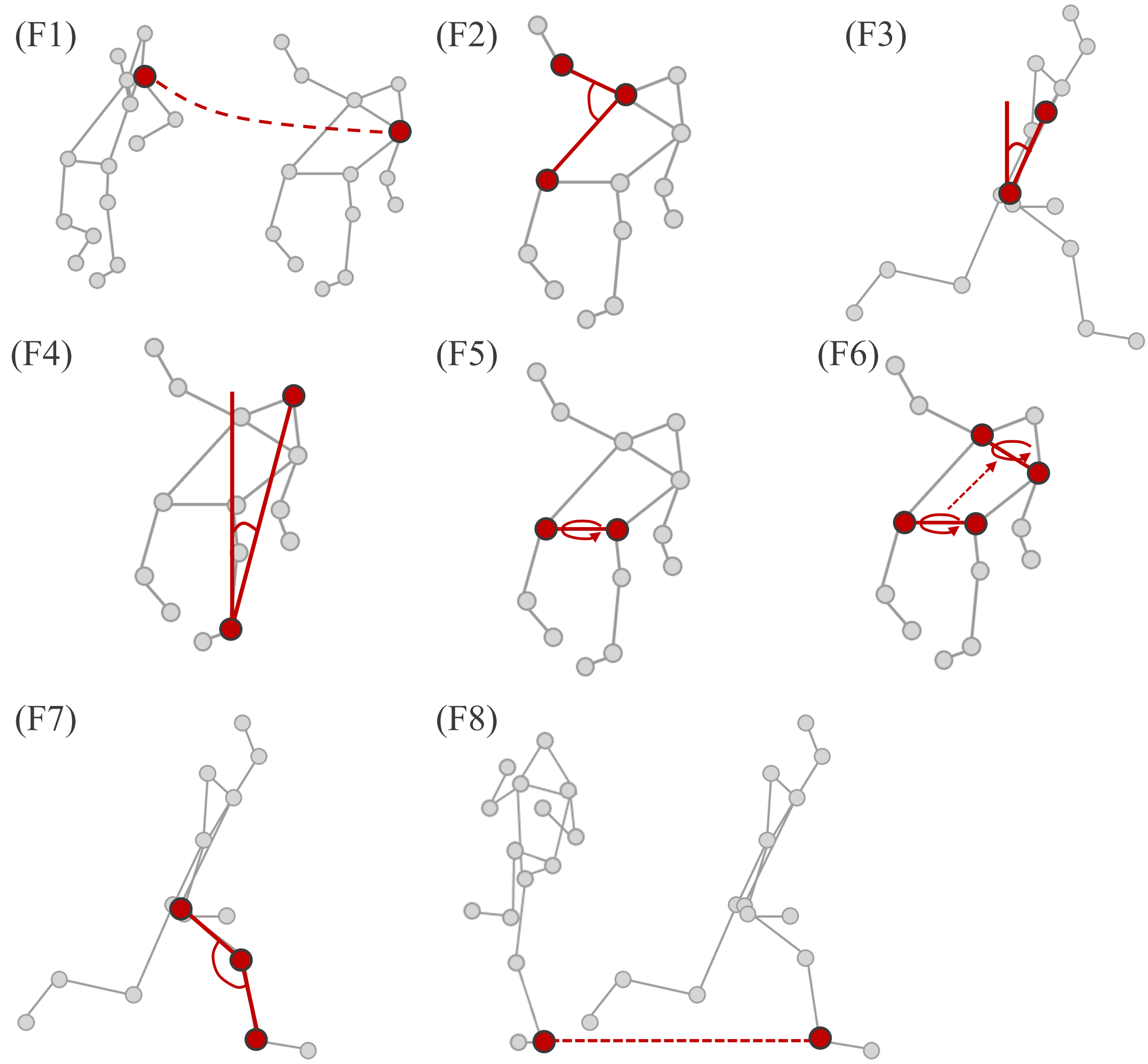

**Fig. 4 Eight biomechanical features used for the verification of the generated motion** (F1) shoulder joint movement, (F2) shoulder abduction, (F3) forward trunk tilt, (F4) horizontal trunk tilt, (F5) maximum trunk rotational velocity, (F6) hip-shoulder delay, (F7) knee extension, (F8) stride length

## 3. Results

Figure 5 shows an example of the motion sequences reconstructed using the transformer-VAE. The average reconstruction error per joint per time frame, calculated as RMSE, was 3.2 cm. Figure 6 presents a visualization of the learned latent space using a t-SNE plot [37]. As illustrated, the Transformer-VAE effectively distinguished the motion patterns of individual pitchers. These results suggest that the latent space learned through the PMGF framework sufficiently captures the spatiotemporal structure of the input pitching motions, enabling accurate reconstruction by the decoder.

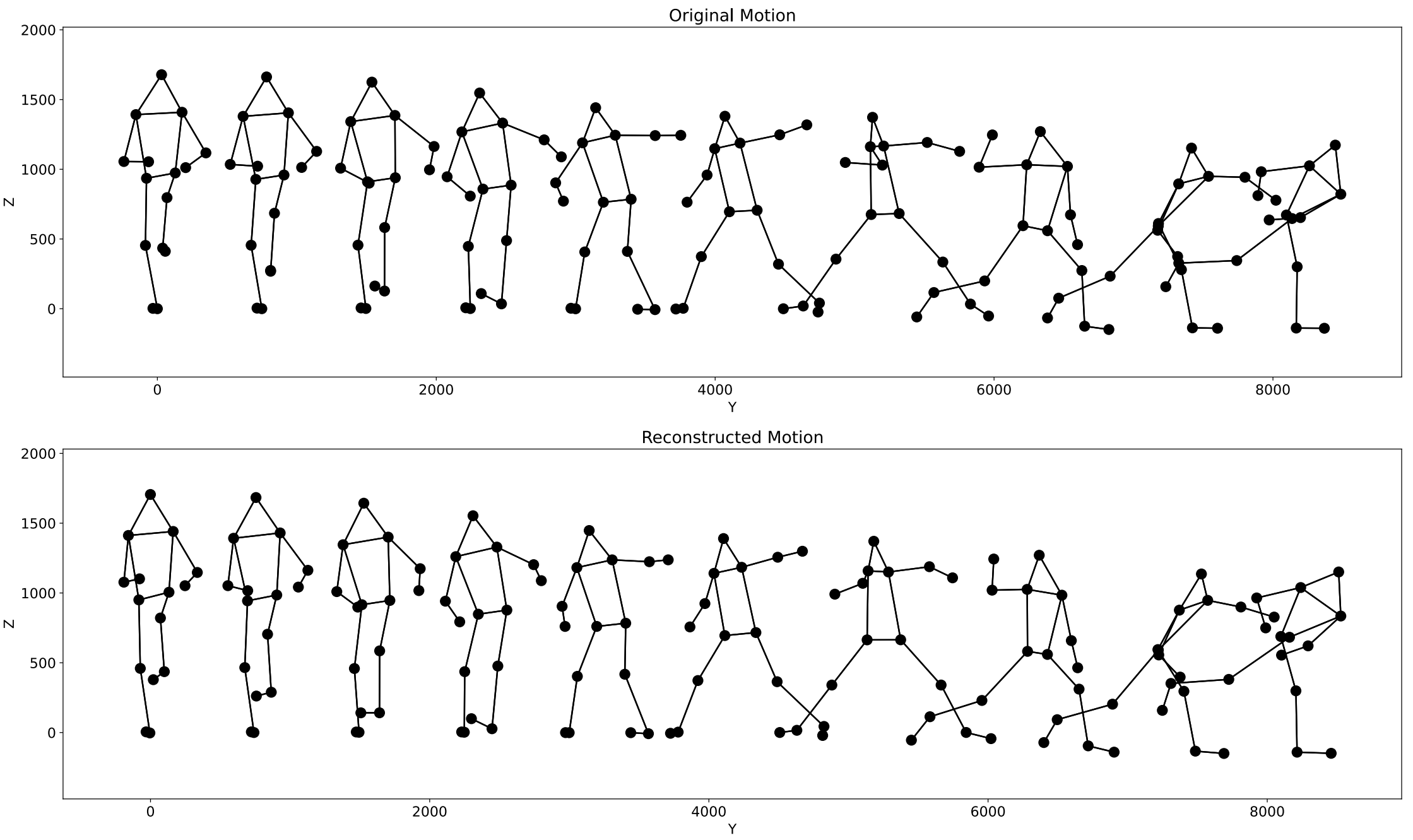

**Fig. 5 Stick picture of original and reconstruction motion sequences**

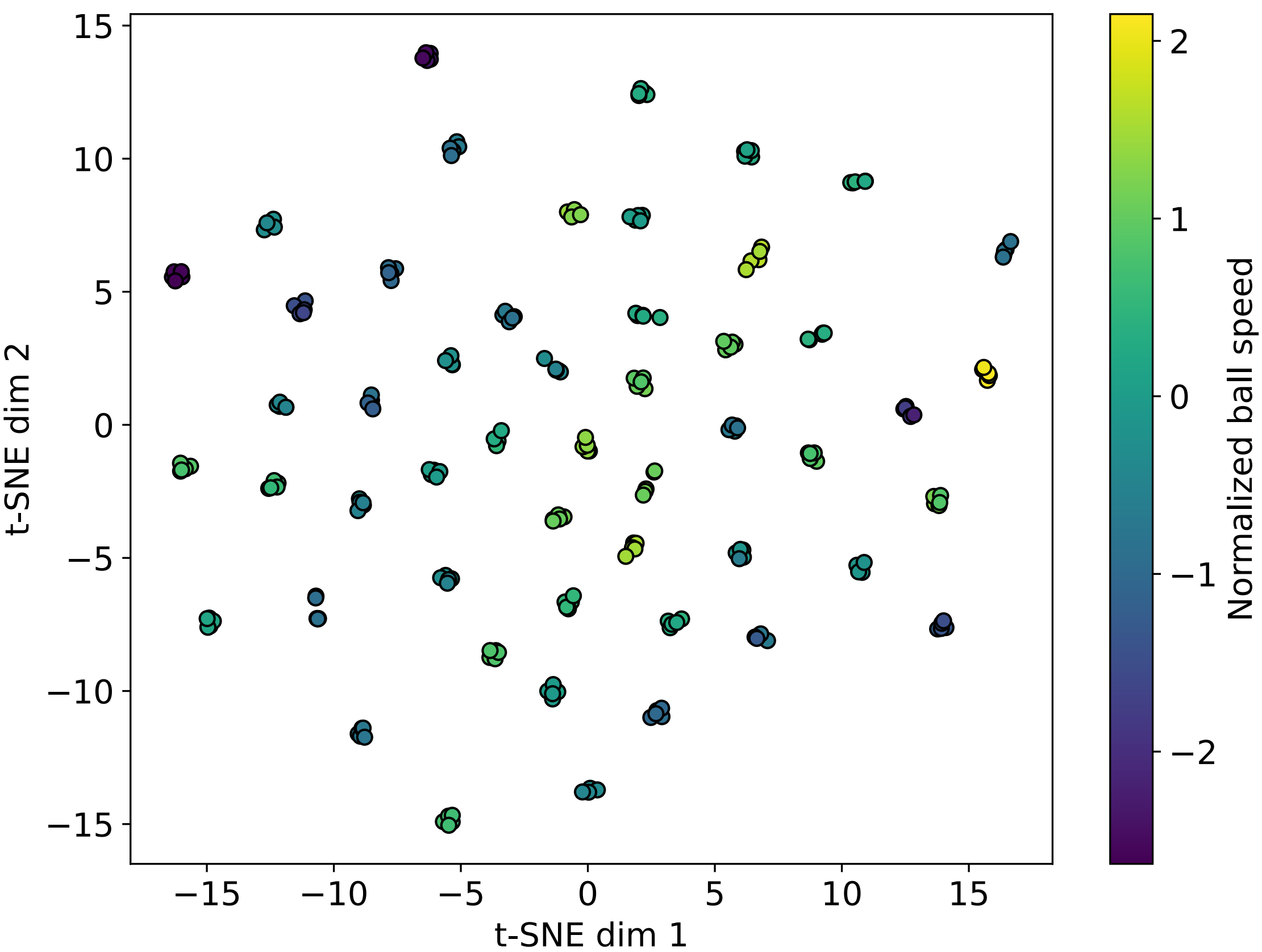

**Fig. 6 t-SNE plot of the learned latent space for five pitches of 51 pitchers**

### 3.1 Motion style transfer between individuals

Figure 7 shows an example of motion style transfer between individuals (from a high school pitcher to an industrial pitcher). The intermediate motions in the second and third rows reflect a blend of characteristics from both the original and target motion patterns. Figure 8 illustrates the DTW distances between the generated motions and both the original and target motions at each transition step (Figure 8a), and shows their means and standard deviations across all pairs (Figure 8b). As shown in the figures, the manipulations performed by PMGF resulted in smooth, continuous transitions in motion patterns across all 1,275 source–target pitcher pairs, demonstrating the effectiveness of latent-space interpolation. Although detailed discussions on this manipulation, including practical applications and potential improvements, will be presented in a later section, this result highlights PMGF’s potential to generate personalized motion guidance by gradually aligning the learner’s movement patterns with ideal ones, such as those of expert athletes or the learner’s own best-performance motions.

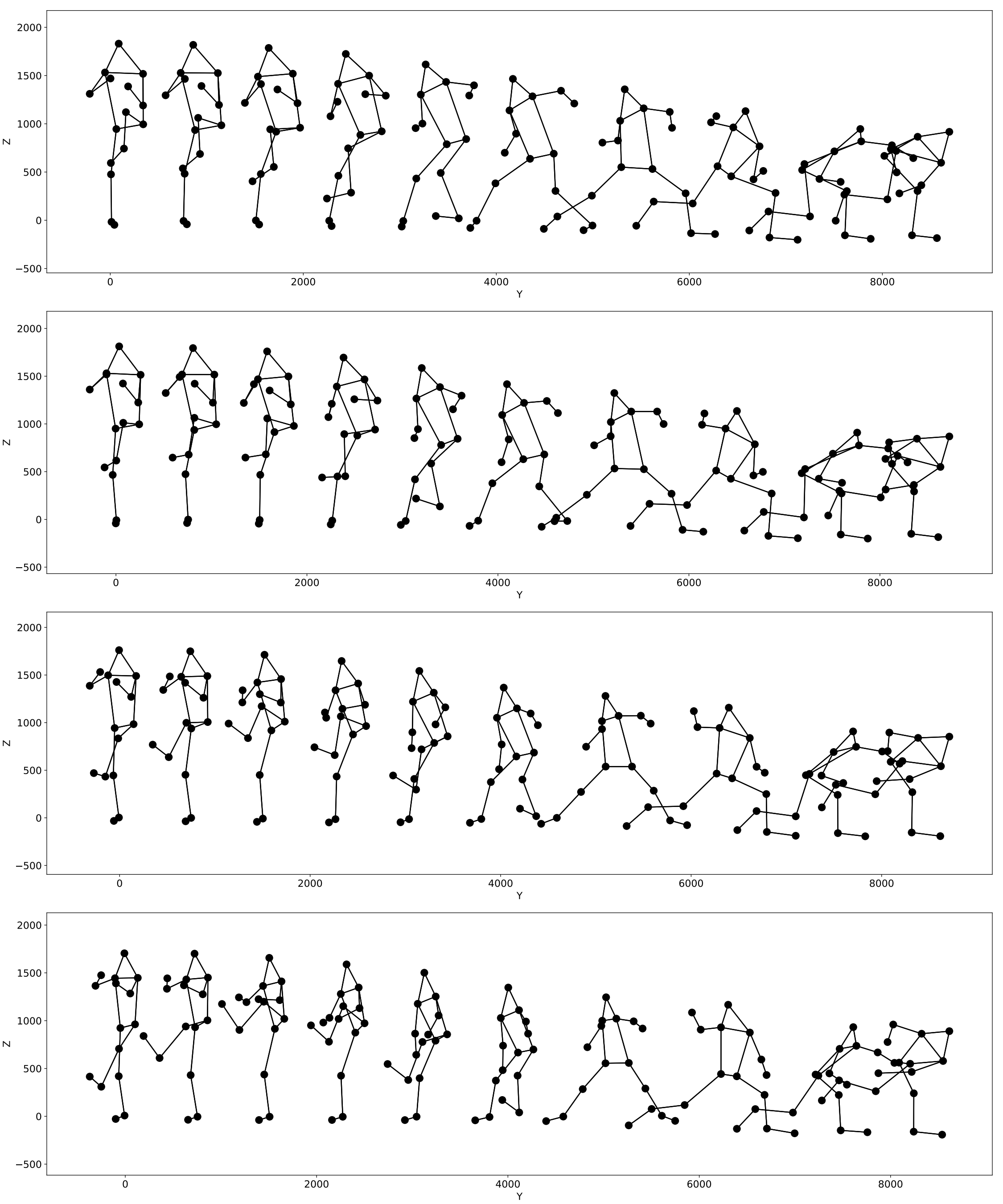

**Fig. 7 Example of motion style transfer between individuals—from high school pitcher to an industrial pitcher**

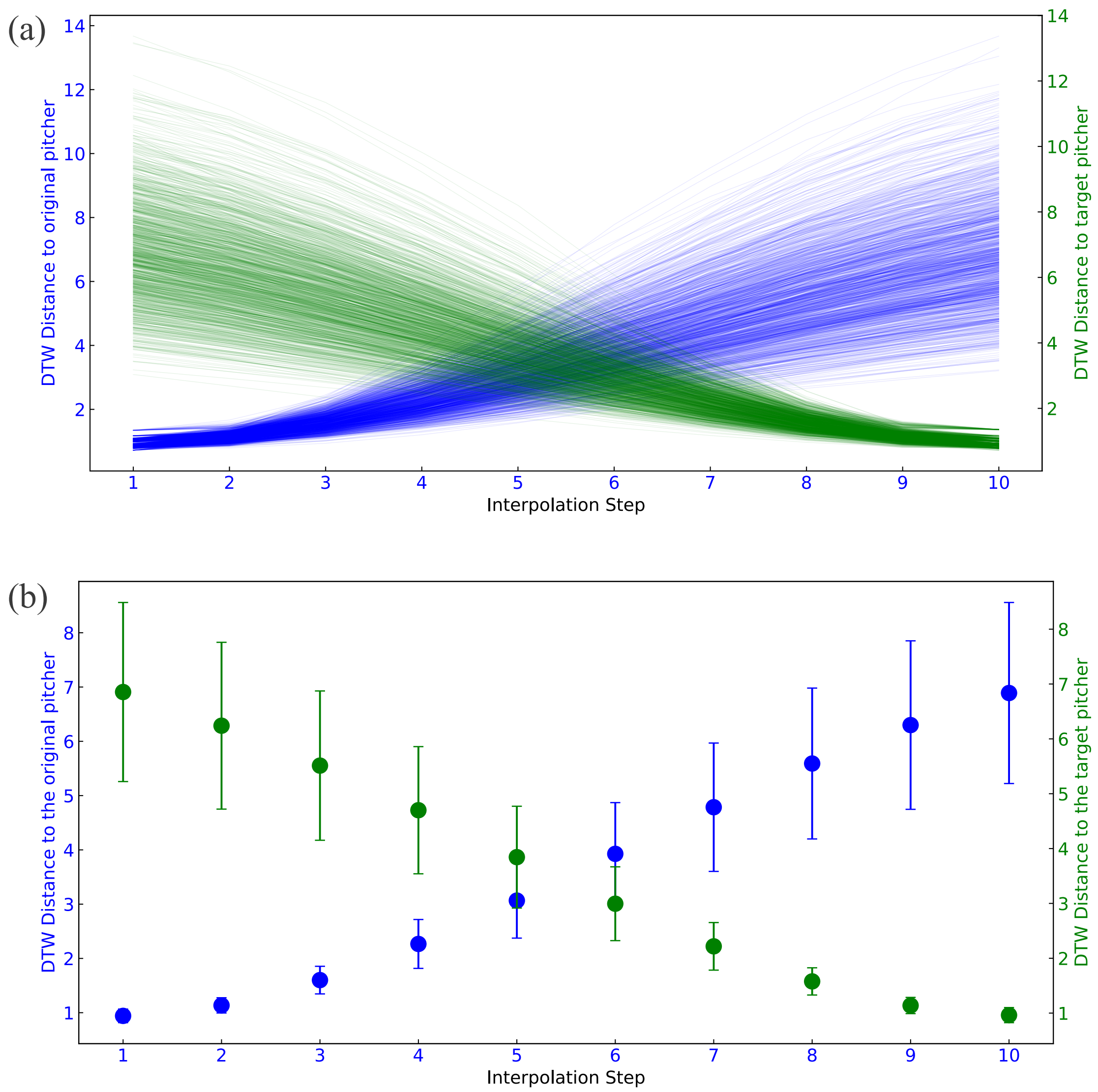

**Fig. 8 Results of DTW analysis for evaluating the smoothness and continuity of the motion pattern transfer** (a) DTW distances between the generated motions and both the original and target motions at each transition step, (b) their means and standard deviations across all pairs. DTW, Dynamic Time Warping

### 3.2 Identifying and shifting the optimal motion to maximize key biomechanical features

Table 1 summarizes the results of biomechanical analyses conducted on the optimized motions generated by the PMGF. Table 2 presents the relationship between the hypersphere radius $r$ and the total effect size. The tables present outcomes from five independent evaluations. In the tables, positive effect sizes indicate a shift in the direction typically associated with higher ball velocity. Figure 9 provides an illustrative example of the motion-pattern changes induced by the manipulation.

As shown in these results, PMGF manipulation consistently produced positive effects on lower-body mechanics. Specifically, significant increases in stride length and knee extension were observed in all runs. Lateral trunk tilt and shoulder abduction also exhibited medium-to-large effect sizes, with significant increases observed in four of five runs and one of five runs, respectively. On average, small effect sizes were observed for shoulder joint movement, forward trunk tilt, and hip–shoulder delay, on average. As shown in Table 2, the total effect size increased with larger values of $r$. The greater standard deviations observed at $r$=4 and $r$=5 is likely due to an increase in the unexplored areas on the hypersphere surface, as the number of optimization iterations was fixed despite the expansion of the search area.

These results suggest that PMGF manipulation tends to shift latent representations in directions that enhance the key kinematic features observed in high-performing pitchers, particularly those related to the lower extremities and trunk posture. This finding supports the potential of PMGF to generate motion guidance that refines an athlete's individual movement patterns to improve performance. One exception was maximum trunk rotational velocity, which consistently exhibited a negative average effect. This negative average effect may be attributed to a limitation of the VAE architecture: its assumption of Gaussian latent distributions may smooth out sharp kinematic features such as peaks in velocity and acceleration, leading to reduced representation of such 'edge' information [38].

**Table 1.** Biomechanical analysis comparing original and optimized pitching motions on hypersphere surface ($r$ = 3)

| | Shoulder joint movement (mm) | | | Shoulder abduction (deg) | | |
|---|---|---|---|---|---|---|
| Model | Mean difference | Adjusted *p* value | Cohen's *d* | Mean difference | Adjusted *p* value | Cohen's *d* |
| 1 | -12.84 | n.s. | 0.33 | 1.47 | n.s. | 0.43 |
| 2 | -10.59 | n.s. | 0.25 | 3.05 | $p < .01$ | 0.81 |
| 3 | -14.38 | n.s. | 0.33 | 2.32 | n.s. | 0.61 |
| 4 | -13.98 | n.s. | 0.34 | 2.61 | n.s. | 0.65 |
| 5 | -11.67 | n.s. | 0.34 | 2.61 | n.s. | 0.69 |

| Model | Forward trunk tilt (deg) | | | Lateral trunk tilt (deg) | | |
|---|---|---|---|---|---|---|
| | Mean difference | Adjusted $p$ value | Cohen's $d$ | Mean difference | Adjusted $p$ value | Cohen's $d$ |
| 1 | 0.09 | n.s. | 0.02 | 0.96 | $p < .01$ | 1.84 |
| 2 | 1.23 | n.s. | 0.23 | 0.30 | n.s. | 0.27 |
| 3 | 0.20 | n.s. | 0.04 | 0.75 | $p < .01$ | 0.91 |
| 4 | 0.44 | n.s. | 0.08 | 1.01 | $p < .01$ | 1.60 |
| 5 | 0.82 | n.s. | 0.17 | 0.70 | $p < .01$ | 1.02 |
| | Maximum trunk rotational velocity (deg/s) | | | Hip-shoulder delay (ms) | | |
| Model | Mean difference | Adjusted $p$ value | Cohen's $d$ | Mean difference | Adjusted $p$ value | Cohen's $d$ |
| 1 | -2.49 | n.s. | -0.03 | 9.78 | n.s. | 0.47 |
| 2 | -6.40 | n.s. | -0.08 | 9.64 | n.s. | 0.44 |
| 3 | -11.5 | n.s. | -0.16 | 6.85 | n.s. | 0.34 |
| 4 | -17.23 | n.s. | -0.24 | 12.16 | n.s. | 0.46 |
| 5 | -8.74 | n.s. | -0.13 | 8.95 | n.s. | 0.41 |
| | Knee extension (deg) | | | Stride length (mm) | | |
| Model | Mean difference | Adjusted $p$ | Cohen's $d$ | Mean difference | Adjusted $p$ | Cohen's $d$ |

| | | value | | | value | |
|---|---|---|---|---|---|---|
| 1 | 5.58 | $p < .01$ | 1.03 | 34.19 | $p < .01$ | 2.25 |
| 2 | 5.50 | $p < .01$ | 0.91 | 37.26 | $p < .01$ | 2.10 |
| 3 | 5.40 | $p < .01$ | 0.89 | 30.35 | $p < .01$ | 2.01 |
| 4 | 5.47 | $p < .01$ | 1.00 | 27.51 | $p < .01$ | 0.82 |
| 5 | 6.41 | $p < .01$ | 1.01 | 33.92 | $p < .01$ | 2.00 |

Table 2. Relationship between hypersphere radius $r$ and total effect size (SD: standard deviation).

| | Total effect size (Cohen's $d$) | | | | |
|---|---|---|---|---|---|
| | $r = 1$ | $r = 2$ | $r = 3$ | $r = 4$ | $r = 5$ |
| Model1 | 2.93 | 5.31 | 6.34 | 4.43 | 7.47 |
| Model2 | 1.94 | 3.70 | 4.92 | 5.95 | 5.29 |
| Model3 | 3.43 | 3.99 | 4.98 | 6.81 | 4.94 |
| Model4 | 2.41 | 3.82 | 4.71 | 5.11 | 5.22 |
| Model5 | 3.64 | 4.42 | 5.48 | 7.86 | 8.42 |
| Mean | 2.87 | 4.25 | 5.28 | 6.03 | 6.27 |
| SD | 0.70 | 0.65 | 0.65 | 1.36 | 1.57 |

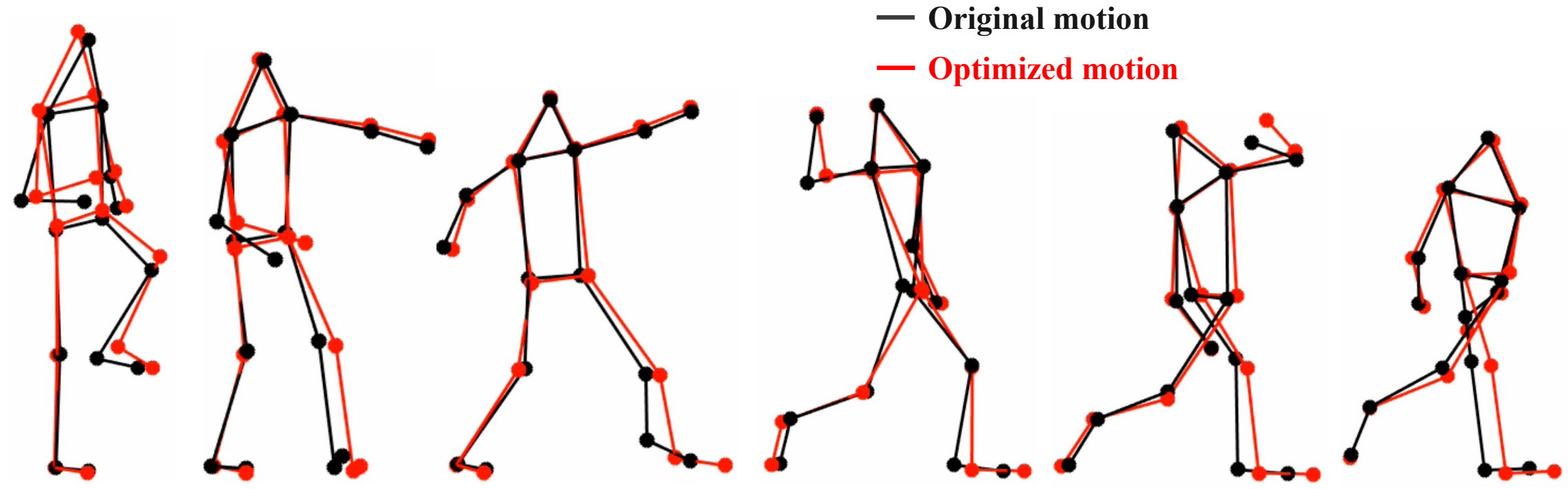

**Fig. 9 An example of motion pattern changes induced by the motion optimization**

## 4. Discussion

The results of the verification experiment using baseball pitchers suggest that the PMGF generates (1) intermediate motion sequences that continuously transition between a learner's current motion and a target (e.g., expert) motion, and (2) biomechanically informed refinements that optimize the learner's motion to enhance key biomechanical features associated with performance indexes such as ball velocity. In the following sections, we discuss the practical and pedagogical implications of these two manipulation strategies, their limitations, and directions for future extension to provide more realistic and versatile guidance across diverse sports contexts.

### 4.1 Practical and pedagogical implications of the PMGF manipulations

Interpolated motion generation between individuals using the PMGF can enhance the effectiveness of observational learning. As discussed in previous sections, the motions generated by the PMGF function as a combined guide that integrates the benefits of both self-and expert modeling [11]. For instance, as illustrated in Figure 10, overlaying the generated motion onto the learner's original motion can provide customized visual feedback that preserves individual movement patterns while subtly guiding the learner toward the target form. In contrast to previous combined-feedback approaches, such as directly overlaying an unaltered expert motion on the learner's performance [39], the PMGF generates guide motions that are more personalized and thus help avoid learning inefficiencies caused by excessive dissimilarity between learner and expert. Moreover, increased self-similarity in the guide motion may facilitate automatic imitation processes toward desired movement patterns [40]. Furthermore, by varying the target motion or introducing stochastic sampling within the latent space, the PMGF may facilitate variability in acquired movement patterns, which is highly emphasized in motor learning studies [41]. From another perspective, using a learner's own best-performing motion as the interpolation target rather than that of another individual may help athletes sustain their best performance.

Whereas the first manipulation aims to align the entire motion pattern with that of a target individual, the second manipulation—shifting the motion pattern in an optimal direction on the hypersphere—offers a more direct and specific refinement of the features associated with a given performance index.

The motion generated by this manipulation is expected to provide effective visual feedback for observational learning. A key advantage of this approach is its broad applicability to different motor skills and performance metrics. Although this study focused on ball velocity, the PMGF could, in principle, be extended to generate personalized guidance for optimizing other performance metrics, such as increasing spin rate or reducing ball-arrival errors, given that relevant biomechanical knowledge is available. Moreover, this strategy can be extended to other motor skills, such as enhancing swing speed in hitting or increasing the running speed of track and field athletes, while preserving unique motion characteristics unrelated to the performance index. Some biomechanical factors that increase with manipulation have also been associated with greater physical stress on the body [5,17]. Therefore, in practical applications, the judgment of human coaches is essential.

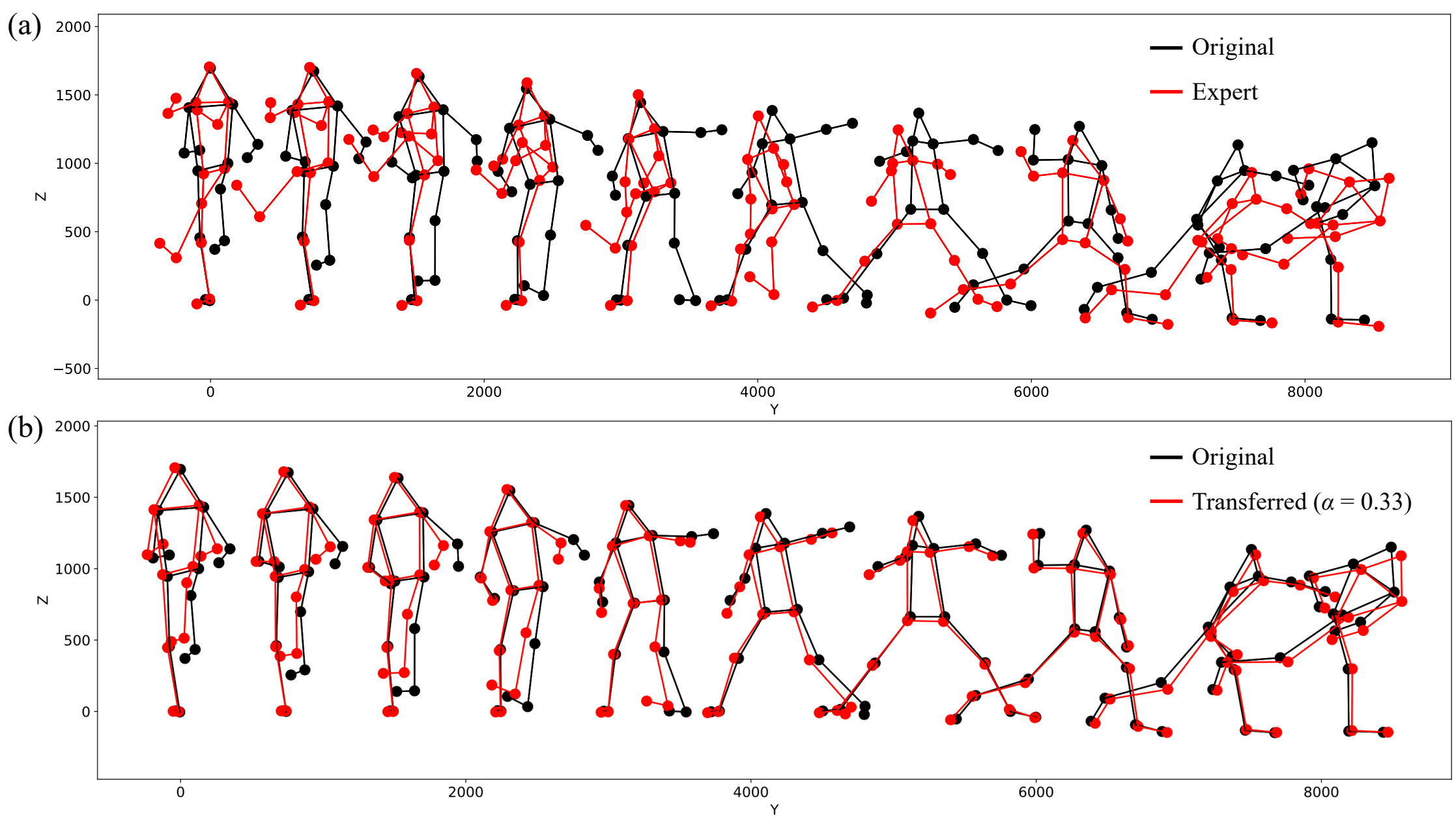

**Fig. 10 A comparison of overlay visualizations for observational learning**: (a) a direct overlay of the learner and the expert pitcher; (b) the generated motion overlaid onto the learner's original motion (a high school pitcher, with 33% style transfer toward an industrial league pitcher)

### 4.2 Limitations

The current PMGF framework exhibits several limitations. One central challenge lies in accounting for individual physical characteristics (i.e., embodiment). Since athletes' creative movements emerge from their unique body and environmental interactions [3], generating more personalized guidance may require conditioning the model on individual physical traits, such as skeletal morphology [42,43]. Such considerations of embodiment should also be reflected in the design of the loss function—for example, by assigning higher penalties to movements that are biomechanically infeasible or that require unrealistic muscular effort. In the current PMGF framework, some generated motions may be physically difficult to execute because of the musculoskeletal limitations of real humans. In addition, instances were sometimes observed in which the reconstructed motions exhibited changes in body dimensions, such as height. Similarly, incorporating task and environmental constraints into the model as conditioning parameters represents an important extension. For instance, in the case of a baseball hit, the optimal swing movement is likely to vary depending on the trajectory of the pitched ball.

Another significant limitation is the difficulty faced by the PMGF in generating guidance for top-tier athletes who aim to improve their performance further. This issue stems from a common limitation in current parametric AI models: the out-of-distribution (OOD) problem [44,45]. Within the ball velocity range of the current dataset (68.47–94.17 mph), the PMGF can reliably generate guidance to support performance improvement. However, it cannot be expected to produce plausible guidance for pitchers already performing at the upper bound of this range unless the motions at 95 or 100 mph follow an extrapolatable trend from existing data. Such limitations may pose a fundamental question regarding the use of AI in high-performance sports coaching: How can we guide top athletes beyond the limits of existing data? The extended PMGF must address this issue by assessing the feasibility and usefulness of generated samples that fall beyond the distribution of the training data [46,47].

### 4.3 Future directions

Finally, we outline potential future extensions of the proposed framework, including possible solutions to the limitations discussed above. Figure 11 illustrates the conceptual design of an extended version of the PMGF, referred to here as *general-PMGF*, which aims to broaden its applicability and generalizability. In essence, the general PMGF builds upon the original PMGF by incorporating ideas from constraint theory [48] and the constraint-led approach [3]. According to these theories, optimal movement emerges through self-organization driven by the complex interaction between an individual's body, environment, and task-specific constraints. Since individual (physical), environmental (including interpersonal interactions), and task constraints can be represented in vector, scalar, or linguistic forms, it is theoretically possible to condition motion generation on these multimodal inputs. Accurate modeling of the interaction between multimodal constraints and motion may benefit from including a reconstruction loss not only for the motion data but also for the constraint information itself [49]. If realized, this framework would allow not only more personalized and context-sensitive motion guidance for athletes but also a data-driven understanding of complex expert-level motor skills. From a practical perspective, incorporating reinforcement learning with a human feedback strategy [50], in which expert coaches or athletes evaluate the generated motions to improve the model iteratively, could be beneficial for addressing the OOD problem and augmenting the training dataset.

Although the architecture presented in Figure 11 is theoretically feasible with sufficient data, the current lack of large-scale motion-capture datasets in sports-specific contexts can be a barrier to its implementation. Therefore, to effectively leverage rapidly advancing generative AI technologies in sports performance analysis, it is essential to promote data sharing and create large and diverse datasets [51].

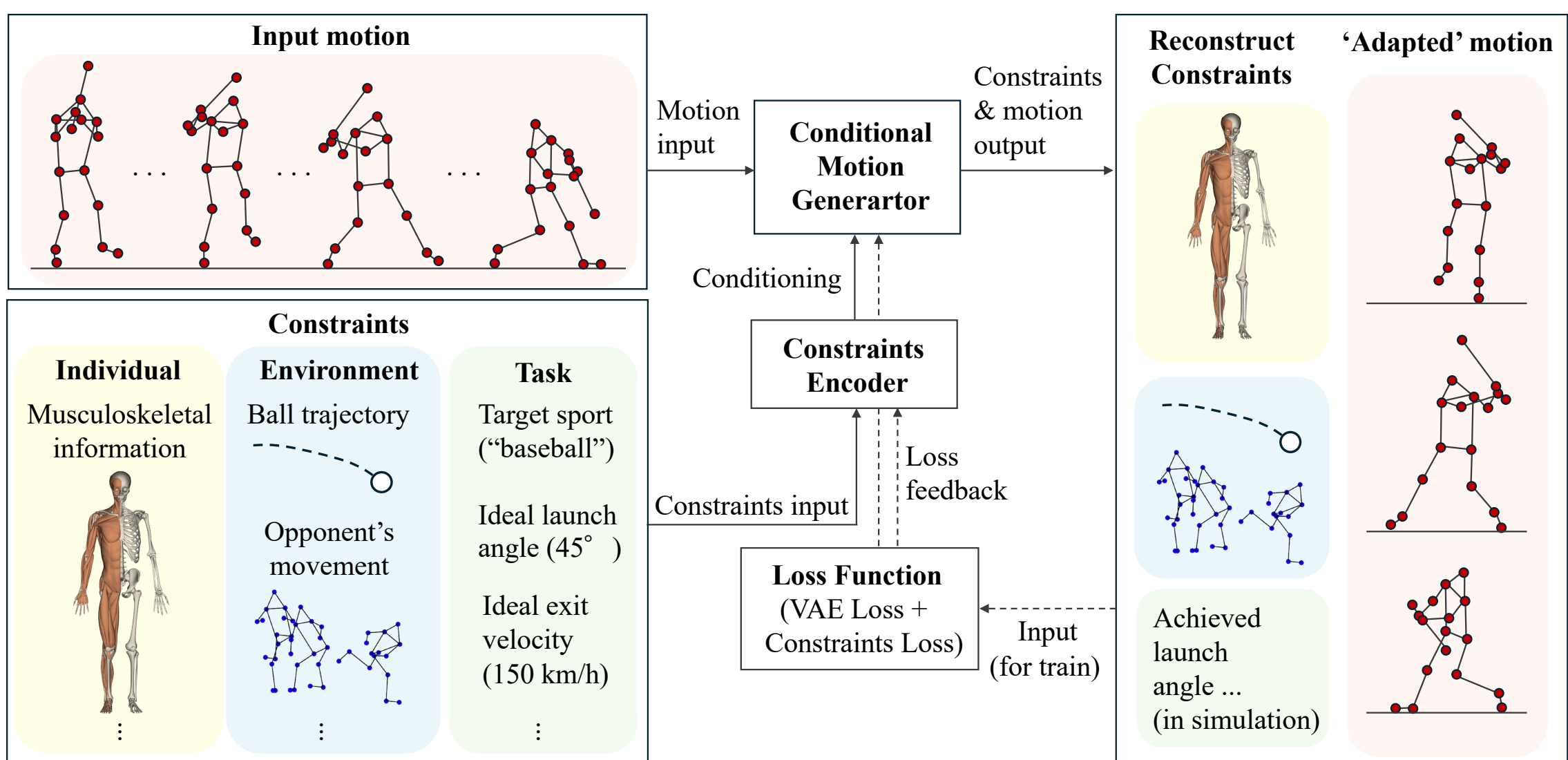

**Fig. 11 A design of the general-PMGF as an extended framework of PMGF.** PMGF, Personalized Motion Guidance Framework

**5. Conclusions**

In conclusion, the two manipulation strategies implemented in the PMGF enable (1) the generation of intermediate motion sequences that smoothly transition between a learner's current motion and a target motion (e.g., expert motion), and (2) biomechanically informed refinements that shift the motion toward optimal directions associated with improvements in key biomechanical features. These outputs can support personalized, athlete-centric coaching practices, such as delivering more individualized visual feedback to enhance observational learning. Although the PMGF has several key limitations that should be addressed in future work, it demonstrates the potential of generative AI as a valuable tool for bridging the gap between the shared characteristics identified through conventional experimental studies and the individualized needs of practical coaching. The development of a more generalized framework integrated with constraint theory may provide more realistic and versatile guidance across diverse sports contexts.

**List of abbreviations**

| | |
|---|---|
| AI | Artificial intelligence |
| DTW | Dynamic Time Warping |
| ES | Evolution Strategy |
| ICCV | International conference on computer vision |
| MSE | Mean squared error |
| OOD | Out-of-distribution |
| PMGF | Personalized Motion Guidance Framework |
| RMSE | Root mean squared error |

**Declarations**

Ethics approval: This study utilized data previously collected for education and training. An opt-out consent framework was adopted in accordance with ethical guidelines. Information about the study, including its purpose and data use, was made publicly available to ensure transparency. Individuals whose data were included had a clear opportunity to decline participation. The study procedures were approved by the Institutional Ethics Committee of the National Institute of Fitness and Sports in Kanoya. All procedures adhered to the principles of the Declaration of Helsinki (approved number: 25-1-25).

Availability of data and material: The source code, anonymized dataset, and analysis scripts used in this study are publicly available on GitHub (https://github.com/takamido/PMGF).

Competing Interests: Ryota Takamido, Chiharu Suzuki and Hiroki Nakamoto declare that they have no competing interests.

Funding: This work was supported by the Japan Society for the Promotion of Science (grant number: JP25K21018).

Authors' contributions: RT was involved in conceptualization, formal analysis, funding acquisition, investigation, original draft preparation, and visualization. CS contributed to methodology, resources, and data curation. HN supervised the study, contributed to methodology, and was involved in project administration. All authors read and approved the final manuscript.

Acknowledgments: This work was supported by the Japan Society for the Promotion of Science (grant number: JP25K21018).